\documentclass[twocolumn,floatfix]{aastex631}
\usepackage{multirow, makecell}
\usepackage{microtype}
\usepackage{graphicx,subfigure}

\usepackage{textcomp}
\begin{document}

\title{UVIT Study of the Magellanic Clouds (U-SMAC). III. Hierarchical Star Formation in the Small Magellanic Cloud Regulated by Turbulence}
\author[0009-0000-3938-162X]{Sipra Hota}
\affiliation{Indian Institute of Astrophysics,
$2^{\rm nd}$ Block, Koramangala,
Bangalore-560034, India}
\affiliation{Pondicherry University,
 R.V. Nagar, Kalapet,
 Puducherry-605014, India\\}

 \author[0000-0002-7203-5996]{Richard de Grijs}
\affiliation{School of Mathematical and Physical Sciences, Macquarie University, Balaclava Road, Sydney, NSW 2109, Australia}
\affiliation{Astrophysics and Space Technologies Research Centre, Macquarie University, Balaclava Road, Sydney, NSW 2109, Australia}
\affiliation{International Space Science Institute---Beijing, 1 Nanertiao, Zhongguancun, Hai Dian District, Beijing 100190, China}
\author[0000-0003-4612-620X]{Annapurni Subramaniam}
\affiliation{Indian Institute of Astrophysics,
$2^{\rm nd}$ Block, Koramangala,
Bangalore-560034, India}

\begin{abstract}
The Small Magellanic Cloud (SMC), a satellite galaxy of the Milky Way, is an irregular dwarf galaxy exhibiting evidence of recent and ongoing star formation. We performed a spatial clustering analysis of far-ultraviolet stars in the SMC younger than 150 Myr using data from the Ultra Violet Imaging Telescope onboard AstroSat. We identified 236 young stellar structures as surface overdensities at different significance levels. The sizes of these structures range from a few parsecs to several hundred parsecs. Their irregular morphologies are characterized by a perimeter--area dimension, derived from the projected boundaries of the young stellar structures, of $D_p = 1.46 \pm 0.4$. The 2D fractal dimensions obtained from, respectively, the number–-size relation and the size distribution are $D_2 = 1.64 \pm 0.03$ and $D_2 = 1.31 \pm 0.16$. These values indicate significant lumpiness among the young stellar structures. In addition, the surface density distribution of the identified structures follows a log-normal distribution. These features are strikingly similar to those of the turbulent interstellar medium, thus supporting the scenario of hierarchical star formation regulated by supersonic turbulence.
\end{abstract}


\section{Introduction}
\label{sec:Introduction}

Stars may be observed as individual field stars or in groups, such as clusters, associations, and even distributed within entire complexes on galaxy-wide scales. It is widely accepted that stars form when gas and dust within molecular clouds undergo gravitational collapse \citep{2003..Lada..SF,2003..Larson..SF,2004...Duch.SF}. The interstellar medium (ISM) is known to exhibit a hierarchical, self-similar structure that spans a range of spatial scales \citep[e.g.,][]{1993..Elmegreen,1996..Elmegreen..Falgarone}. This means that large structures in the ISM, such as giant molecular clouds (GMCs), are composed of smaller, denser structures, which themselves break down into even tinier, denser formations, and so forth. This spatial distribution in the ISM is mainly controlled by turbulence, magnetic fields, and self-gravity \citep{1996..Elmegreen..Efremov}. Young stars with ages $\leq$100 Myr show similar hierarchical patterns as the ISM gas \citep[e.g.,][]{2015..Gouliermis..HSF..CT,2018...Sun..SMC,2022MNRAS..Miller..LMC}. Observations and simulations both suggest that young stellar structures form following the approximately fractal arrangement of the ISM \citep{1996..Elmegreen..Falgarone,2001..Elmegreen..&..Elmegreen,2017..Vazquez}. This process is known as `hierarchical,' `scale-free,' or `fractal' star formation \citep{1993..Gomez,1995..Larson,1998...Bate}. 

The Small Magellanic Cloud (SMC) is an irregular, gas-rich dwarf satellite galaxy of the Milky Way (MW), located at a distance of $\sim$60 kpc \citep{2000..Cioni..distance,2015..de..Grijs,2020ApJ...904...13Graczyk..SMC..dist}. Its close proximity, resolved stellar populations, and low-metallicity environment render the SMC a unique astrophysical laboratory for studies of galaxy formation and evolution. The SMC has likely experienced multiple collisions with its larger companion galaxy, the Large Magellanic Cloud \citep[LMC;][]{2009..Harris..SFH..MCs,2010..Glatt..MCs..Interaction,2016..Besla..multi_LMC_SMC..int,2018..Rubele..MCs..int,2022...SD..Massana,2024..Dhanush}. The most recent interaction, which occurred some 150--300 Myr ago, resulted in the formation of the Magellanic Bridge, which connects the LMC and the SMC and is composed of gas, and stars, \citep{1963..Hindman..MB,1985..Irwin..MB,1990AJ.Irwin..MB,1994...Gardiner,2007..Muller_MBR,2019Zevick..MB}, and also of other tidal structures formed recently \citep[e.g., counter bridge, western halo;][]{2014..Dais,2017..Ripepi..NE-SW..SMC,2018..Zivic..MCs..PM,2021..Dais..counter_bridge}. Using proper-motion measurements from the {\sl Hubble Space Telescope}, \citet{2006..Kallivayali..firts_passage_MCs} found that, contrary to prior expectations of having completed multiple orbits, the Clouds are more likely on their first infall trajectory toward the MW \citep{2007..Besla..first..passage..MCs, 2009..Bekki..Chiba}. However, we note that \citet{2023..Vasiliev} proposed a scenario in which the Clouds are currently on their second passage. Past interactions among the MW, LMC, and SMC have led to the formation of the Magellanic Stream and its counterpart, the Leading Arm \citep{1992..Liu..MS,1994...Gardiner,1998..mw-lmc-smc-inter,2007..Besla..first..passage..MCs,2017..Belokurov..Magellanic..system,2020..Luchin..MS}. As a result, these interactions have perturbed the SMC's dynamical evolution \citep{1998Natur.394..752Putman..Magellanic..Stream,2020..Massana..SMC..int,Tatton..2021..inter}.
    
\citet[][]{1985..Hodge..Associations..SMC} identified 70 probable stellar associations in the SMC. Although the average size of these associations (77 pc) is comparable to that of associations in the LMC \citep{1970...Lucke..LMC..associations}, it is smaller than that of the typical stellar association found in the MW, Messier 31 (M31), or Messier 33 (M33). \citet[][]{1991..Battinelli..OB..SMC} proposed a unique method for detecting associations in the SMC based on the observed spatial distribution of OB stars and their projected separations. He found that the mean size of the detected OB associations was 90 pc, closely matching the result reported by \citet{1985..Hodge..Associations..SMC}. In a study of hierarchical star formation in M33, \citet{2007...Bastian..HSF..M33} found that the range of characteristic sizes might vary depending on the selection criteria applied, i.e. the mean size can be different based on the `breaking scale' adopted and the minimum star-count level implemented in the identification algorithm. \citet{2008..Gieles..Stellar..Str..SMC} investigated stellar populations in the SMC with ages ranging from 10 Myr to 1 Gyr, using $V$ vs. $V-I$ color--magnitude diagrams (CMDs) from the Magellanic Clouds Photometric Survey \citep[MCPS;][]{2002..Zaritsky..MCPS...MCPS}. By applying the two-point correlation function and a method based on minimum spanning trees, they found that the stellar distribution shows a high degree of substructure at young ages ($\sim$10 Myr), which disappears on a timescale of $\sim$75 Myr. \citet{2010..Bonatto..HSF..L&SMC} used cluster and non-cluster data from the \citet{2008..Bica..Catalog} catalog of the Magellanic Clouds and found that the distribution of young clusters exhibits a high degree of spatial self-correlation. It particularly correlates with the distribution of star-forming structures, a pattern not observed for older clusters. They observed that the structures' size distribution follows a power law, which aligns with expectations linked to the hierarchical star-formation scenario. However, since the \citet{2010..Bonatto..HSF..L&SMC} study relied entirely on the extended catalog of \citet{2008..Bica..Catalog}, assessing the impact of selection biases and incompleteness is difficult. In their recent study of hierarchical star formation in the SMC, \citet{2018...Sun..SMC} used near-infrared data from the VISTA survey of the Magellanic Clouds \citep[VMC;][]{2011..VMC..survey..M-R-L..Cioni} and found that young stellar structures exhibit irregular morphologies, which however follow power-law distributions in both mass and size. Their findings also suggest that hierarchical star formation may be controlled by supersonic turbulence within the ISM.

To investigate recent star formation, which is expected to exhibit a hierarchical pattern, access to the far-ultraviolet (FUV) wavelength range is crucial, since it serves as a prominent tracer of the youngest stars, which show a peak in their flux distribution at FUV wavelengths \citep{1997..Cornett..UIT,2023..UVIT..SMC..Devaraj,2024..Hota..SMC..Shell,2024Hota..FUV..catalog..SMC}. In this study, we use the FUV catalog of the SMC derived by \citet[][hereafter H24]{2024Hota..FUV..catalog..SMC} and employ a contour-based map analysis technique \citep{2015..Gouliermis..HSF..CT,2017..Gouliermis..HSF..CT} to identify young stellar structures as overdensities at various significance levels from the surface density map of young stars. We consider all complexes, associations, and clusters, which we collectively refer to as young stellar structures, without distinguishing among them. The objective of the present paper is to investigate the hierarchical pattern of young massive stars in the SMC using FUV observations, with the aim of understanding the clustering, overall structure, and dynamics of the youngest stellar population on galaxy-wide scales. The paper is structured as follows. In Section~\ref{sec:dm} we describe the data used and the methods employed to identify young stellar structures. Section~\ref{sec:CDP} presents the correlation and distribution of various defining parameters such as the number of stars within the structures, size, perimeter, area, and surface density. In Section~\ref{sec:dis}, we discuss the results from the application of a range of constraints and provide comparisons with other studies. Finally, we conclude the paper in Section~\ref{sec:cs}.

\section{Data and Methodology}
\label{sec:dm}

\subsection{Ultraviolet Data}
\label{subsec:data}

\begin{figure}
    \centering
    \includegraphics[width=\columnwidth,trim={1.2cm 0.5cm 2.2cm 1.5cm},clip]{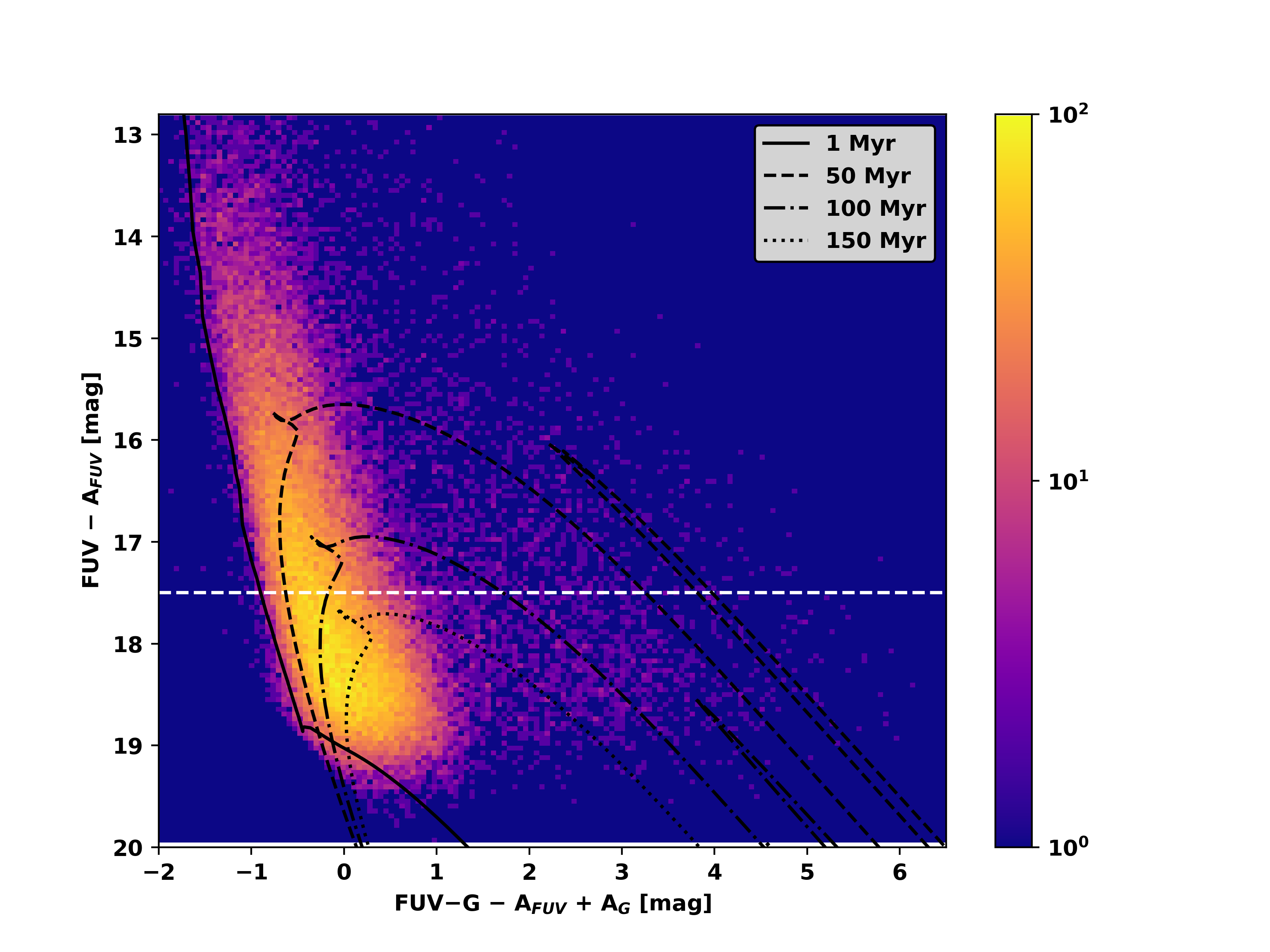}
    \caption{FUV--optical color--magnitude Hess diagram of the most probable FUV stars ($\sim$62,900) in the SMC. The color bar represents the number of stars in each color--magnitude bin. The white dashed line represents our FUV magnitude cutoff at 17.5 mag.}
    \label{fig:FUV-optical_cmd}
\end{figure}

In our study, we use the ultraviolet (UV) point-source catalog of the SMC compiled by H24 (see their Table 2), obtained from 39 fields across the SMC observed by the Ultra Violet Imaging Telescope \citep[UVIT;][]{2017June..TondonUVIT} onboard AstroSat. UVIT is a twin telescope with a diameter of 38 cm, allowing simultaneous observations in FUV (130--180 nm), near-UV (NUV; 200–-300 nm), and visible (VIS; 350–-550 nm) wavebands. The VIS observations are used primarily for drift correction resulting from the spacecraft's motion \citep{2017June..TondonUVIT}. UVIT has a field of view of $\sim$0.5 degree diameter with a spatial resolution of $\sim$1.4$''$. Detailed information regarding the instrument and its calibration is provided by \citet{2011PostemaDetectorPC,2012kumarUVIT,10.1117/12.2235271,2017Sep..TondonUVIT}. The SMC UVIT catalog contains $\sim$76,800 FUV sources that cover the SMC's main body and inner Wing region, with sources detected up to 20 mag in the F172M filter \citep[silica, $1717\pm125$\AA;][]{2020TondonUVITfilters}, with a maximum error of 0.2 mag (signal-to-noise ratio, $\frac{S}{N}$ $\geq$ 5).

In this study, we have considered only the $\sim$62,900 most probable SMC FUV stars (see the final column of H24's Table 2) with {\sl Gaia} Data Release 3 (DR3) counterparts that meet the following criteria: (i) Renormalised Unit Weight Error (RUWE) $< 1.4$ \citep[][]{2018..Lindegren}, (ii) a classification probability greater than 31\%, which indicates the likelihood that a source is a member of the SMC, and (iii) $G < 19.5$ mag \citep[criteria (ii) and (iii) were obtained from][]{2023..Gaia_catalog_SMC_Prob_Jimenez}. These criteria were adopted to exclude foreground stars. Detailed information as regards foreground decontamination can be found in Section 2.5 of H24.

Since our identification of young structures depends on their surface density, a high completeness level of our sample of FUV stars is essential. H24 performed an FUV completeness check of their catalog by conducting artificial-star tests (ASTs). They achieved a recovery rate of approximately 90\% for stars with FUV magnitudes up to 18 mag, increasing to 100\% completeness at 16 mag (refer to their Figure 3b). In our analysis, we considered FUV stars with magnitudes down to 17.5 mag ($N \simeq 20,800$), where the completeness is greater than 90\%. We verified the results using different magnitude cut-offs from 16.0 to 18.0 mag, in steps of 0.5 mag: see Section~\ref{sec:dis}. The mass range spans from 5 to 40 $M_\odot$, corresponding to FUV magnitudes between 18 and 13.3 mag (photometric mass; H24).

\autoref{fig:FUV-optical_cmd} shows the FUV–-optical CMD for about 62,900 SMC FUV stars. The FUV and $G$ magnitudes have been corrected for extinction using an average color excess of $E(B-V) = 0.05$ mag \citep{2011AJ....141..158H,2021..Skowron..Reddening}, adopting the extinction laws of \citet{1989...Cardelli} and \citet{1994..O.Dennell..extinction..law}, and $R_V= 3.1$ \citep{2008..Girardi..extinction..law,2024...Goradon..extinction}. The Padova group's PARSEC isochrones \citep[v2.0;][]{2012...Bressan..Padova..PARSEC..iso} for ages of 1, 50, 100, and 150 Myr, assuming a distance modulus of $m-M = 18.96$ mag \citep{2015..de..Grijs} and a metallicity of $Z = 0.002$ \citep[][]{1992ApJ...metallicity,2008A&A...metallicity,2017A&A...metallicity} are overlaid on the FUV--optical CMD. We note that FUV stars brighter than 18 mag are younger than 150 Myr, and these stars are primarily O- and B-type main-sequence stars, as well as giants (H24). Therefore, the stars selected for this study with FUV magnitudes brighter than 17.5 mag are younger than 150 Myr.

\subsection{Surface density distribution}
\label{subsec:kde}

\begin{figure}[ht]
    \centering
    \includegraphics[width=\columnwidth,trim={0.6cm 0.2cm 1.5cm 1.2cm},clip]{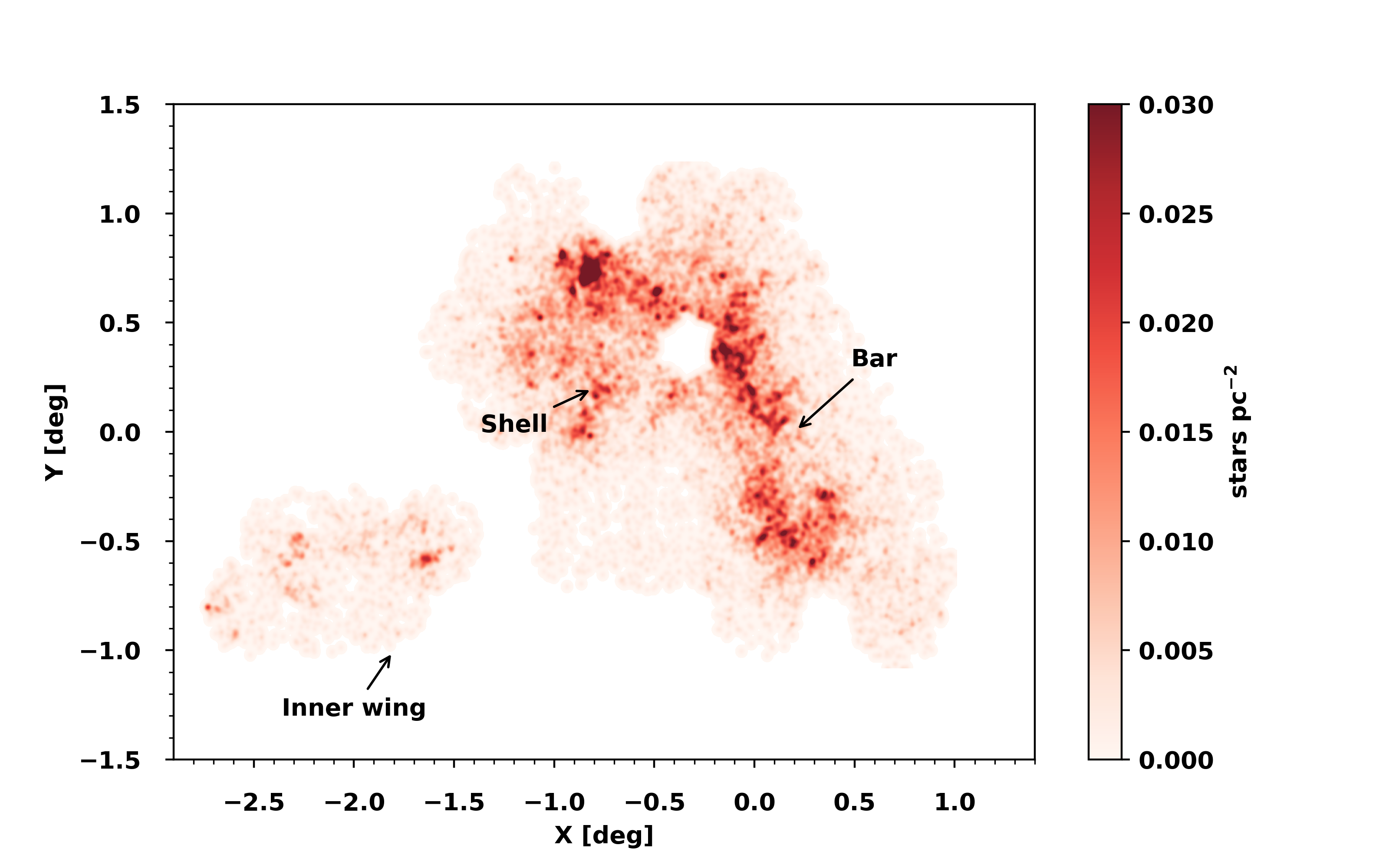}
    \caption{Surface density map (kernel density estimation) of the SMC. The color bar represents the number of stars pc$^{-2}$. The bar, shell, and inner wing of the SMC are marked with black arrows. XY are the projected coordinates, with the optical center at $\alpha_{\text{SMC}} = 00^{\text{h}}52^{\text{m}}12^{\text{s}}.5$, $\delta_{\text{SMC}} = -72^\circ49'43''$ \citep[J2000;][]{1972..de..Vaucouleure..optical..center}.}
    \label{fig:surface_density}
\end{figure}

The SMC's spherical coordinates were projected onto the $XY$ plane using the zenithal equidistant projection method, adopting the $X$ and $Y$ conversions given by \citet{2001AJ..van..der..Marel..&..Cioni..SMC..projection}. The optical center of the SMC was taken at $\alpha_{\mathrm{SMC}} = 00^{\mathrm{h}} 52^{\mathrm{m}} 12^{\mathrm{s}}.5$ and $\delta_{\mathrm{SMC}} = -72^{\circ} 49' 43''$ \citep[J2000;][]{1972..de..Vaucouleure..optical..center}. To analyze the surface density distribution of our selected sample (FUV magnitudes $<$ 17.5 mag), we employ kernel density estimation (KDE). This technique involves convolving the stellar distribution with a Gaussian kernel to create a 2D binned representation of the selected sample. The width of the kernel is chosen based on the scale we are interested in. A narrow kernel can show fine details but may also introduce noise, whereas a wider kernel can reduce noise but may lose some detail. We tested kernel widths from 5 to 20 pc and determined that a width of 10 pc provides a suitable balance between reducing noise and losing detail. The same kernel size has also been used in previous studies of the Magellanic Clouds \citep{2017..Sun..LMC..Bar,2017...Sun..30,2018...Sun..SMC,2022MNRAS..Miller..LMC}. In the KDE map, there is a gap at (${\mathrm{\alpha}}$,${\mathrm{\delta}}$) $\equiv$ ($14^{\circ}$, $-72.5^{\circ}$) where UVIT data are unavailable (H24).

From the KDE map of the selected sample (age $\leq$ 150 Myr; \autoref{fig:surface_density}), we note the broken bar, which extends in the northeast--southwest direction across the SMC \citep[H24;][]{2013..Sewilo..bar..wing,2019Dalal..MORPHOLOGY..smc}, a shell-like structure \citep[][H24]{2001..Maragoudaki..SMC..Morphology}, and the inner SMC Wing \citep{2000..Zaritsky..SMC..morphology,2000...Cioni..SMC..morphology,2013..Sewilo..bar..wing,2023..Oliveira..SMC..wing..mb}. The overall surface density distribution of the selected SMC sample is irregular and clumpy \citep[e.g.,][H24]{2000..Zaritsky..SMC..morphology,2000...Cioni..SMC..morphology, 2013..Sewilo..bar..wing,2019Dalal..MORPHOLOGY..smc,2024..Hota..SMC..Shell}. The median, mean, and standard deviation of the KDE surface density are 0.00008 pc$^{-2}$, 0.002 pc$^{-2}$, and 0.005 pc$^{-2}$, respectively. These values are consistent with those found by \citet{2018...Sun..SMC} and \citet{2022MNRAS..Miller..LMC} for the SMC and LMC, respectively, using VMC data. Our identification and analysis of young structures is based on this surface density map. 

\subsection{Detection of Young Stellar Structures and their parameters}
\label{subsec:YSS}

\begin{figure*}[ht]
    \centering
    \includegraphics[width=\linewidth]{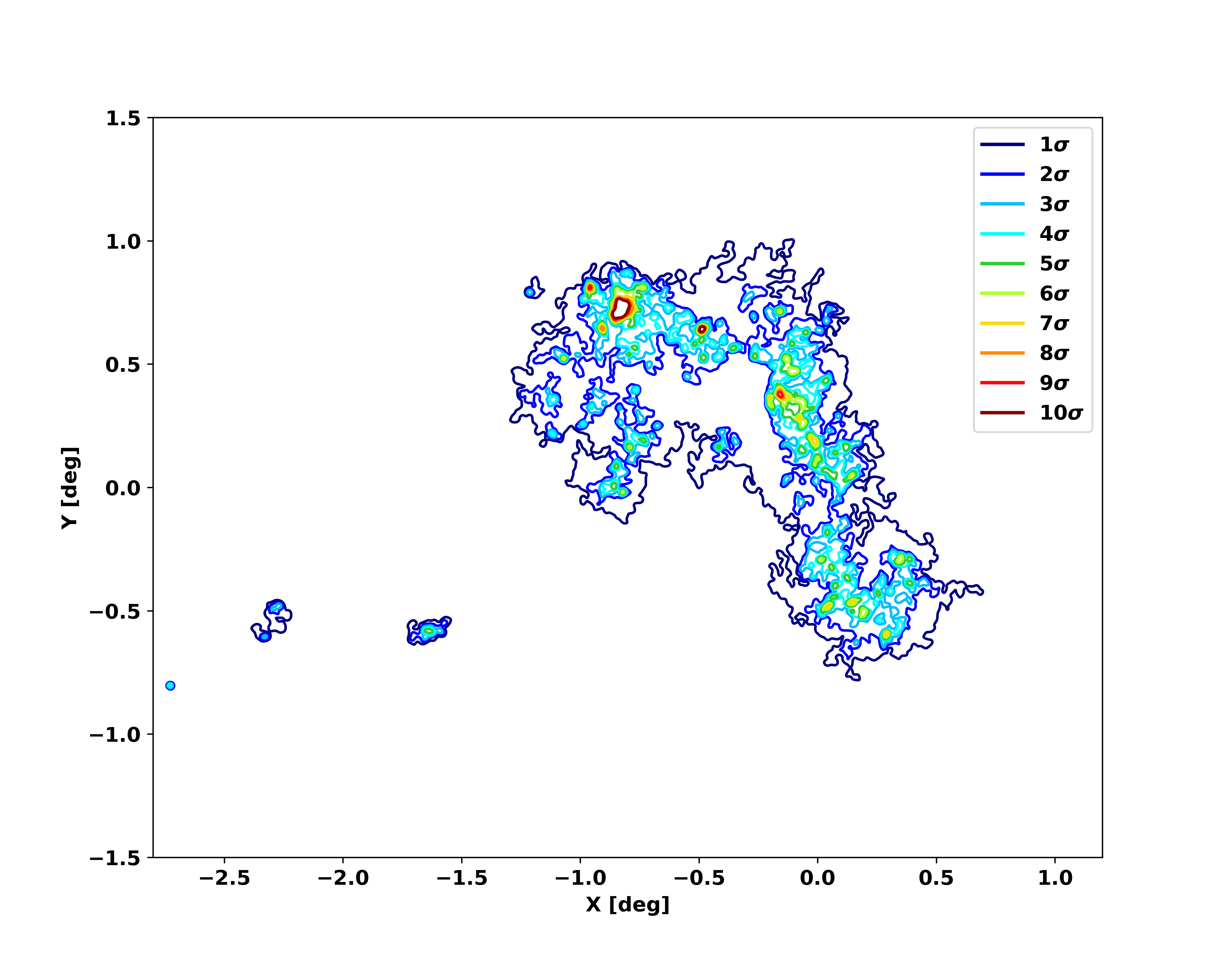}
    \caption{Detected young stellar structures colored by their significance levels.}
    \label{fig:contour_smc}
\end{figure*}

To identify young stellar structures in the SMC's surface density map (\autoref{fig:surface_density}), we apply a contour map-based technique \citep{2015..Gouliermis..HSF..CT, 2017..Gouliermis..HSF..CT}. This technique uses estimates of surface overdensities across a range of significance levels ($n$ times the standard deviation of the surface density from the KDE map, $\sigma$, where $n$ is an integer) to detect young stellar structures. In our work, a range of significance levels, from 1$\sigma$ + the median value of the KDE density to 10$\sigma$ + the median value of the KDE density, in equal steps of 1$\sigma$ (where $\sigma = 0.005$ pc$^{-2}$) was used to obtain the isodensity contours in the KDE map. Note that the median value of the KDE map is significantly smaller than the standard deviation in all less populous regions. At each significance level, any isodensity contour enclosing an overdensity is identified as a candidate young stellar structure.\footnote{These isodensity contours represent the projected boundaries of the candidate young stellar structures.} Using this method, we have detected approximately 370 young stellar structures of all kinds. 

In the next step, we estimate the physical parameters of the detected young stellar structures. The radius of a stellar structure ($R$) is determined by the radius attained by a circular area equivalent to the projected boundary. The total number of stars within a structure is estimated by the number of stars of the selected sample ($N$) within the boundary. The surface density of a structure is estimated by computing the ratio of the number of stars within its boundary to the area of the structure, using $\Sigma = \frac{N}{\pi R^2}$.

There is a possibility that some of the structures identified may be spurious detections. To address this potential concern, we applied two additional criteria adopted from previous studies \citep{2018...Sun..SMC,2022MNRAS..Miller..LMC}: (1) a genuine structure must contain a minimum number of stars ($N_{\rm min}$), and (2) structures at the 1$\sigma$ and 2$\sigma$ significance levels must enclose the contours of structures at higher significance levels (3$\sigma$ to 10$\sigma$). The choice of $N_{\rm min}$ is arbitrary; a higher value helps to reject most unreliable structures but may also remove genuine ones, whereas a lower value may result in more unreliable detections. Therefore, we set $N_{\rm min}$ = 5, consistent with previous studies, to reduce contamination by spurious structures \citep{2007...Bastian..HSF..M33,2009..Bastian..HSF.LMC,2017..Gouliermis..HSF..CT,2018...Sun..SMC,2022MNRAS..Miller..LMC}. Following application of these two criteria, we retained a total of 236 young stellar structures: see \autoref{fig:contour_smc}. This shows the non-uniform and hierarchical pattern of the detected young stellar structures in the SMC, which is similar to those in many star-forming regions and entire galaxies \citep{2010..Gouliermis..HSF..NGC6822,2015..Gouliermis..HSF..CT,2014..Gusev..HSF..NGC628,2017...Sun..30,2017..Sun..LMC..Bar,2018...Sun..SMC,2022MNRAS..Miller..LMC}. We emphasize the hierarchical pattern of the detected structures. Table~\ref{Tab:Young_parameters} lists the details of individual structures across a range of significance levels; Table~\ref{Tab:avg_parameters} summarizes the demographic data for the detected young stellar structures as a function of significance level.

\begin{deluxetable*}{lcccccccc}
\label{Tab:Young_parameters}
\tabletypesize{\scriptsize}
\tablewidth{0pt}
\tablecaption{Details of the detected young stellar structures.}
\tablehead{
    \colhead{ID} & 
    \colhead{Level ($\sigma$)} & 
    \colhead{$X$ (deg)} & 
    \colhead{$Y$ (deg)} & 
    \colhead{RA (deg; J2000)} & 
    \colhead{Dec (deg; J2000)} & 
    \colhead{$N$} & 
    \colhead{$R$ (pc)} & 
    \colhead{$\Sigma$ (pc$^{-2}$)} 
}
\startdata
    1 & 1 & $-$2.345	&$-$0.591 &  21.35 & $-$73.26 &   31   &  17.06	& 0.034 \\ 
    2 & 1 & $-$2.277    &$-$0.520 &  21.09 & $-$73.20 &   93   &  29.67 & 0.034 \\ 
    3 & 1 & $-$1.643	&$-$0.580 &  18.92 & $-$73.33 &  152   &  34.64 & 0.040 \\ 
    4 & 1 & $-$0.336	&    0.228 &  14.31 & $-$72.59 & 15,424 & 366.07 & 0.037 \\
\enddata
\tablecomments{This table includes the identification number (ID), significance level ($\sigma$), central coordinates ($X$ and $Y$, in degrees), central spherical coordinates ($RA$ and $Dec$, in degrees; J2000), the number of stars ($N$) within each structure, the radius/size ($R$, in parsecs), and the surface density ($\Sigma$, in pc$^{-2}$) for each detected structure (columns 1 to 9). Only the first four records are shown here, as an example. Projected XY coordinates of the SMC are defined with respect to the galaxy's optical center at $\alpha_{\text{SMC}} = 00^{\text{h}}52^{\text{m}}12^{\text{s}}.5$, $\delta_{\text{SMC}} = -72^\circ49'43''$ (J2000; \citealt{1972..de..Vaucouleure..optical..center}).} The full table is available in MRT format.
\end{deluxetable*}

\begin{deluxetable*}{lccccccc}
\label{Tab:avg_parameters}
\tabletypesize{\scriptsize}
\tablewidth{0pt}
\tablecaption{Demographics of the detected young stellar structures at each significance level.}
\tablehead{
    \colhead{Level} & 
    \colhead{$N_{\rm str}$ (No. structures)} & 
    \colhead{$R_{\rm min}$ (pc)} & 
    \colhead{$\langle R \rangle$ (pc)} & 
    \colhead{$R_{\rm median}$ (pc)}&
    \colhead{$R_{\rm max}$ (pc)} & 
    \colhead{$N_{\rm sum}$ (No. stars)} & 
    \colhead{$\langle \Sigma \rangle$ (pc$^{-2}$)} 
}
\startdata
    1 &  5 & 17.06 & 93.17 & 29.67 & 366.07 & 15734  & 0.035 \\ 
    2 & 19 &  8.33 & 35.14 & 19.00 & 175.37 & 10607  & 0.056 \\ 
    3 & 59 &  3.55 & 13.71 &  8.19 &  98.97 &  7275  & 0.083 \\ 
    4 & 60 &  2.46 & 11.52 &  8.49 &  47.96 &  4138  & 0.109 \\ 
    5 & 47 &  3.07 &  8.76 &  6.14 &  37.86 &  2258  & 0.140 \\ 
    6 & 24 &  2.34 &  8.23 &  7.59 &  31.51 &  1170  & 0.180 \\ 
    7 & 11 &  2.27 &  7.72 &  5.95 &  26.88 &   626  & 0.222 \\ 
    8 &  5 &  3.63 &  9.82 &  8.13 &  23.68 &   447  & 0.224 \\ 
    9 &  4 &  3.14 &  9.03 &  6.01 &  20.95 &   342  & 0.259 \\ 
    10 & 2 &  5.62 & 12.13 & 12.13 &  18.64 &   267  & 0.269 \\ 
\enddata
\tablecomments{This table presents the significance level ($\sigma$), number of structures ($N_{\rm str}$), minimum size ($R_{\rm min}$), average size ($\langle R \rangle$), median size ($R_{\rm median}$), maximum size ($R_{\rm max}$), number of stars ($N_{\rm sum}$), and the average surface density ($\langle \Sigma \rangle$) of the detected young stellar structures at each significance level (columns 1 to 7).}
\end{deluxetable*}

\section{Correlations and Distributions of Key Parameters}
\label{sec:CDP}

\subsection{Perimeter--Area relation}
\label{subsec:PAR}

\begin{figure}[ht]
    \centering
    \includegraphics[width=\columnwidth,trim={0.4cm 0.1cm 1.5cm 1.2cm},clip]{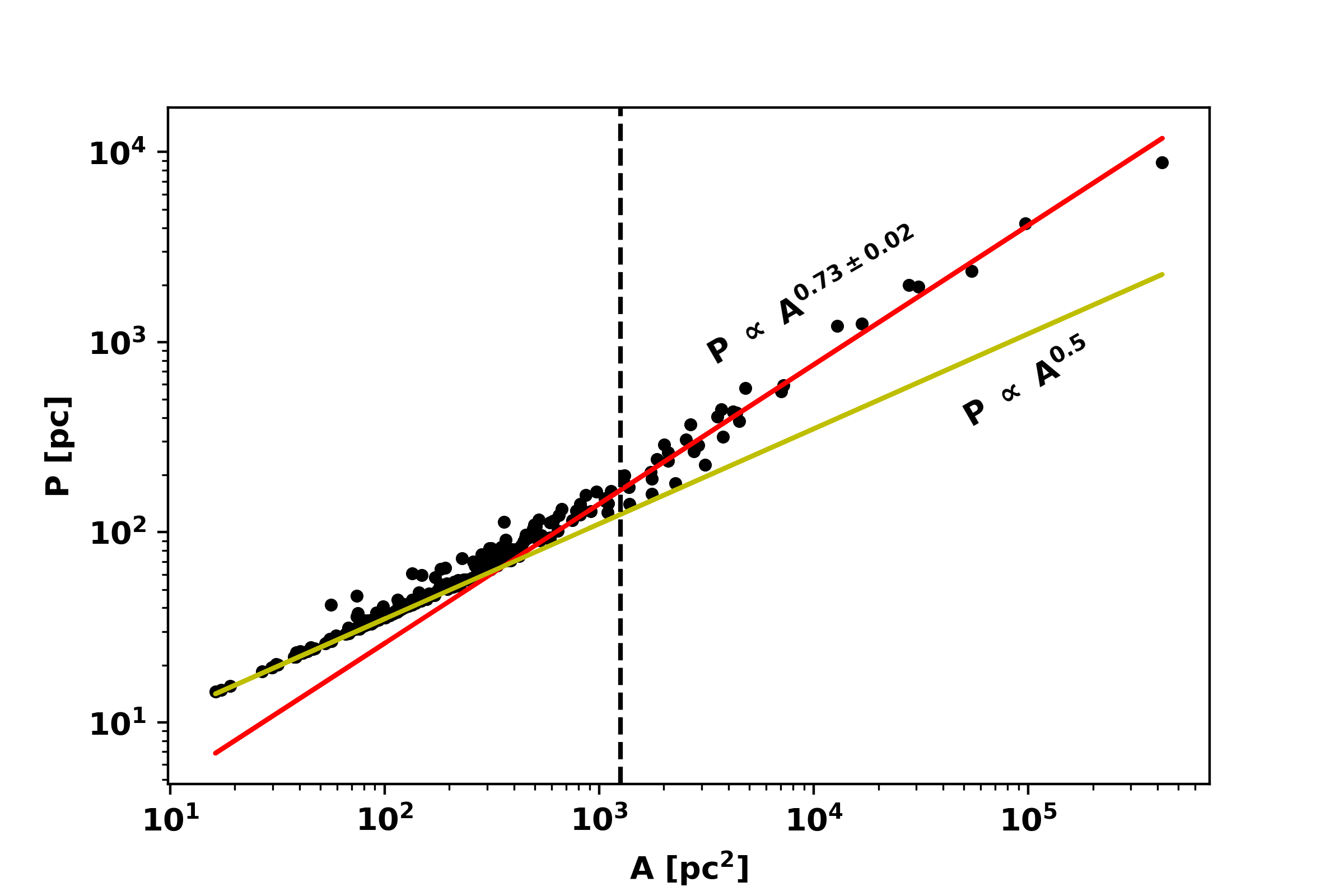}
    \caption{Perimeter vs. area of the detected young stellar structures. The black dashed line 
 represents the area cut-off, $A_{20} \approx 1.3 \times 10^3$ pc$^2$, which corresponds to a size of $R = 20$ pc. For areas beyond this cut-off radius, the data is fitted with a single power-law function (as indicated by the red solid line).}
    \label{fig:P_A_rel}
\end{figure}

From the contour plot (\autoref{fig:contour_smc}), we note that the boundaries of the detected young stellar structures are irregular. These irregularities in the morphology of the detected structures can be quantified by a perimeter--area relation. \autoref{fig:P_A_rel} shows the perimeter ($P$) versus area ($A$) of the boundaries of the detected young stellar structures. Circular structures follow the perimeter--area relationship, $P \propto A^{0.5}$. We consider the area threshold $A_{20}$ at $R = 20$ pc to be greater than the resolution of the KDE map (\autoref{fig:surface_density}), so that resolution effects become negligible. For an area $\leq$ the area at $R = 20$ pc ($A_{20} \approx 1.3 \times 10^3$ pc$^2$), the detected structures follow the circular perimeter--area relation shown by the yellow solid line in \autoref{fig:P_A_rel}. The data points above this cut-off ($> A_{20}$) follow the perimeter--area relation with a power-law slope of $\alpha = 0.73 \pm 0.02$ (the power-law slope was found by means of least-squares fitting) shown as the red line, while the area threshold, $A_{20}$, is represented by the dashed black line. The perimeter--area dimension, ${D_p}$, can be estimated using $P \propto A^{D_p/2}$ \citep{1991..Falgarone..PA}. For the observed young stellar structures, the perimeter--area dimension is $D_p = 2\times \alpha = 1.46 \pm 0.04$. This value is considerably higher than the perimeter--area dimension of a circular (smooth) structure, which therefore quantitatively reflects the irregular shapes of the detected young stellar structures (with R $>$ 20 pc).


\subsection{Number--Size relation}
\label{subsec:MSR}

\begin{figure}[ht]
    \centering
    \includegraphics[width=\columnwidth,trim={0.4cm 0.1cm 1.5cm 1.2cm},clip]{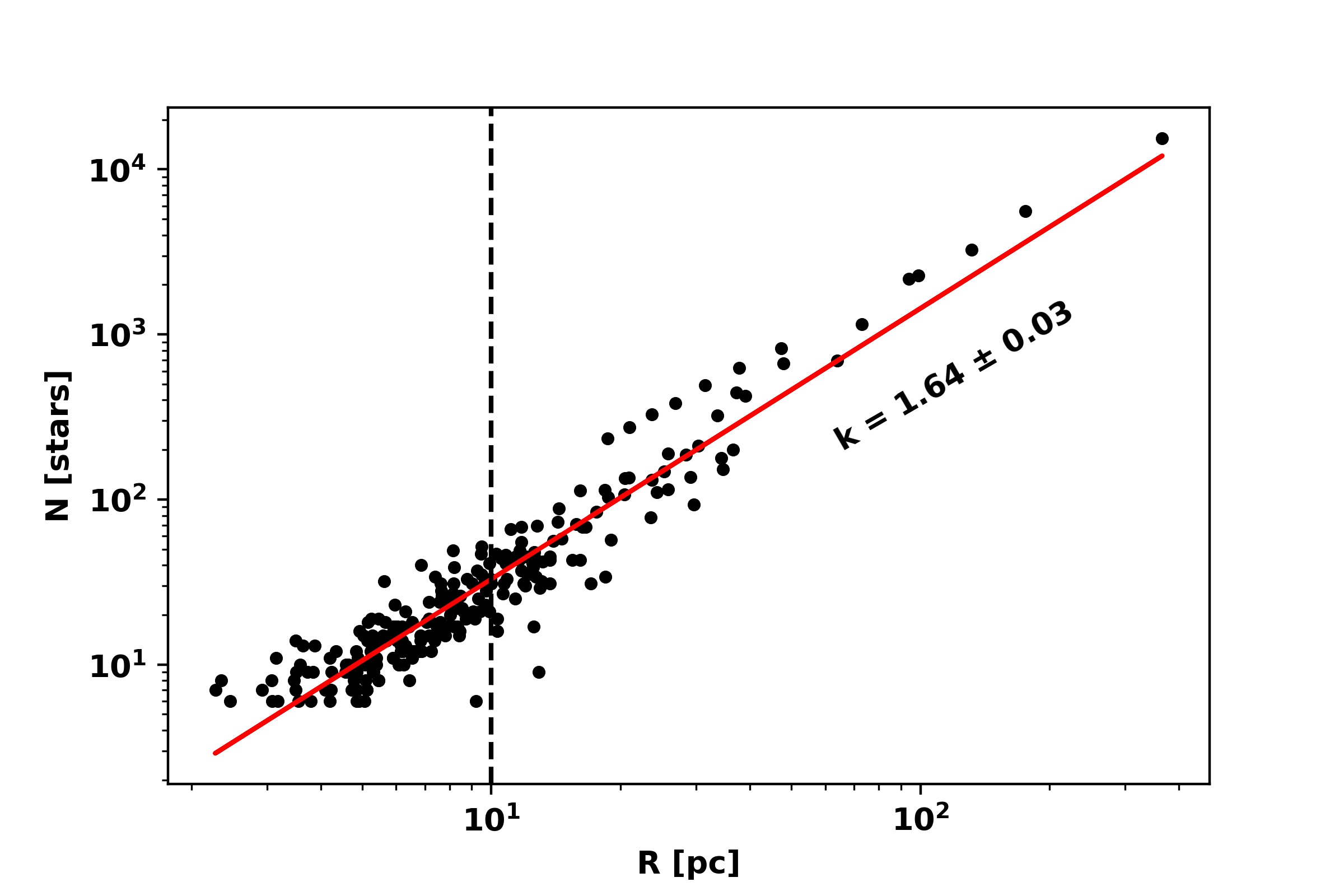}
    \caption{Number--size relation of the detected young stellar structures. The dashed line indicates the spatial resolution of the KDE (see \autoref{fig:surface_density}). The solid red line represents the best single power-law fit.}
    \label{fig:mass_size_rel}
\end{figure}

\begin{figure}[ht]
    \centering
    \includegraphics[width=\columnwidth,trim={0.6cm 0.2cm 1.8cm 1.2cm},clip]{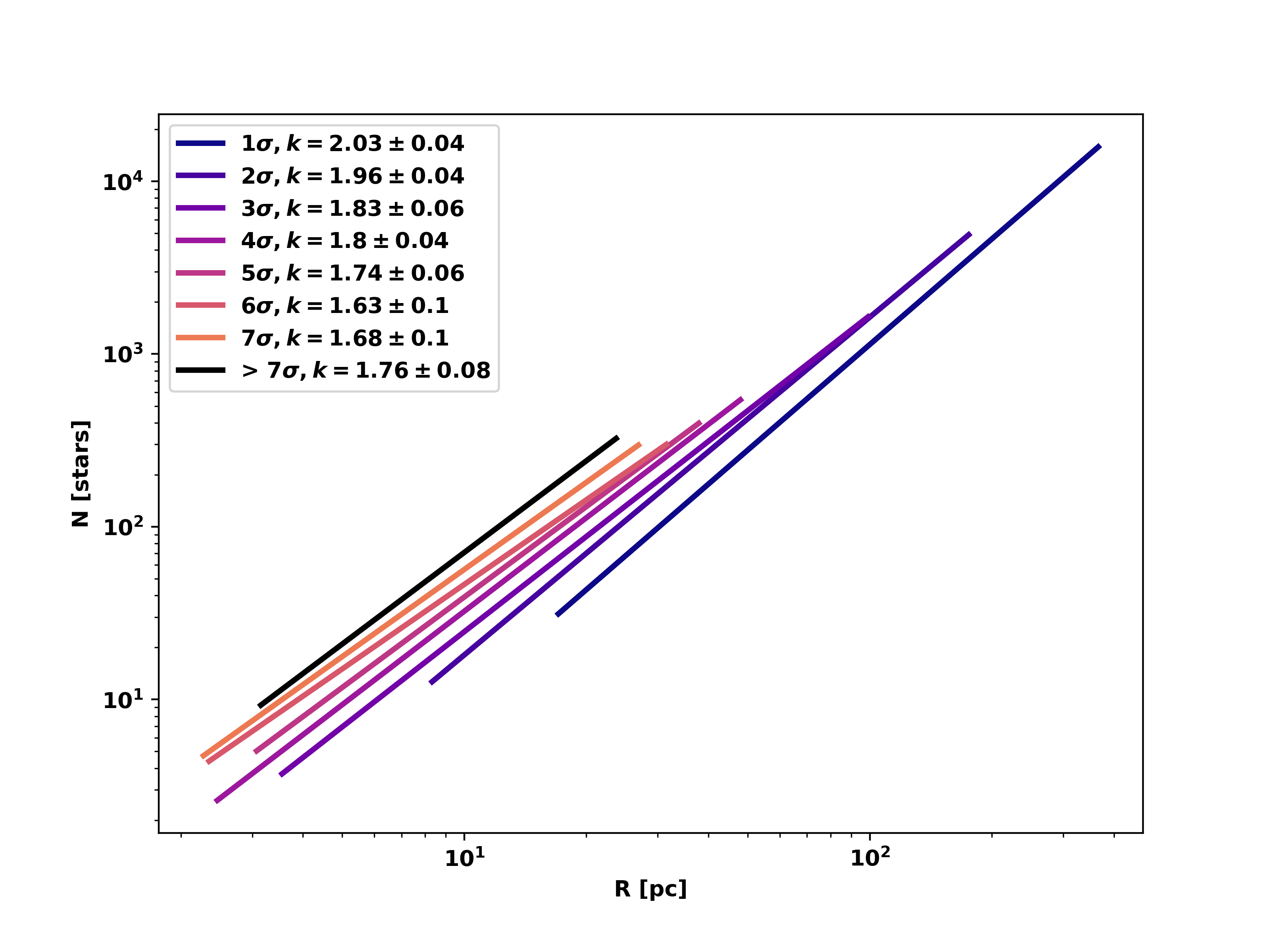}
    \caption{Power-law fits to the relationship between the numbers and sizes of the detected young stellar structures at different significance levels. Since the number of data points at significance levels $>$ 7$\sigma$ is small, we combined all data points for significance levels 8--10$\sigma$.}
    \label{fig:mass_size_rel_diff_level}
\end{figure}

\autoref{fig:mass_size_rel} shows the variation of the numbers of stars within the structures ($N$) as a function of the size of the detected young stellar structures. We estimate the best-fitting slope, using a least-squares fit, of $k = 1.64 \pm 0.03$. While UV observations are biased in favor of detecting young, massive stars, and the stellar initial mass function (IMF) is not fully sampled at this wavelength range, the number--size relation can still serve as a reliable proxy for the mass--size relation. This is because the mass–-size relation of structures estimated using massive stars is expected to exhibit a similar slope to the mass–-size relation derived for the total stellar population within the structures. 

The mass--size relation for masses that are hierarchically clustered follows the relationship $M \propto R^{D_2}$ \citep{1983fractal}\footnote{The 2D fractal dimension, ${D_2}$, is not the same as the perimeter--area dimension, ${D_p}$.}. The projected fractal dimension of the detected young stellar structures is ${D_2} = k = 1.64 \pm 0.03$. We have estimated the 2D fractal dimension by calculating the slopes of least-squares fits to structures at different significance levels: see \autoref{fig:mass_size_rel_diff_level}. In this figure, we display only the power-law fits. Because the number of points is limited to five or fewer, we used a single power law to describe the number--size relation for all structures with significance levels $> 7\sigma$ (i.e. $8\sigma$ to $10\sigma$). Note that the slopes of the best fits, and consequently the fractal dimensions of the structures, vary from ${D_2}$ = 1.6 to 2.0 across significance levels. \citet{1987..Feitzinger..FSF} showed that uniform artificial distributions, such as rhombic and quadratic patterns, exhibit fractal dimensions close to 2. In contrast, star-forming regions typically display lower fractal dimensions, ranging from 1.4 to 1.9. This implies that stellar structures with a fractal dimension near 2 have a uniform surface density, while those with lower fractal dimensions are marked by irregular, patchy distributions. The derived ${D_2}$ values, computed at various significance levels and for the entire sample, suggest that the structures exhibit non-uniform, clumpy distributions.


\subsection{Distributions of size, number, and surface density}
\label{subsec:DSMSD}

\begin{figure}
    \centering
    \includegraphics[width=\columnwidth,trim={0.6cm 0.2cm 1.5cm 1.2cm},clip]{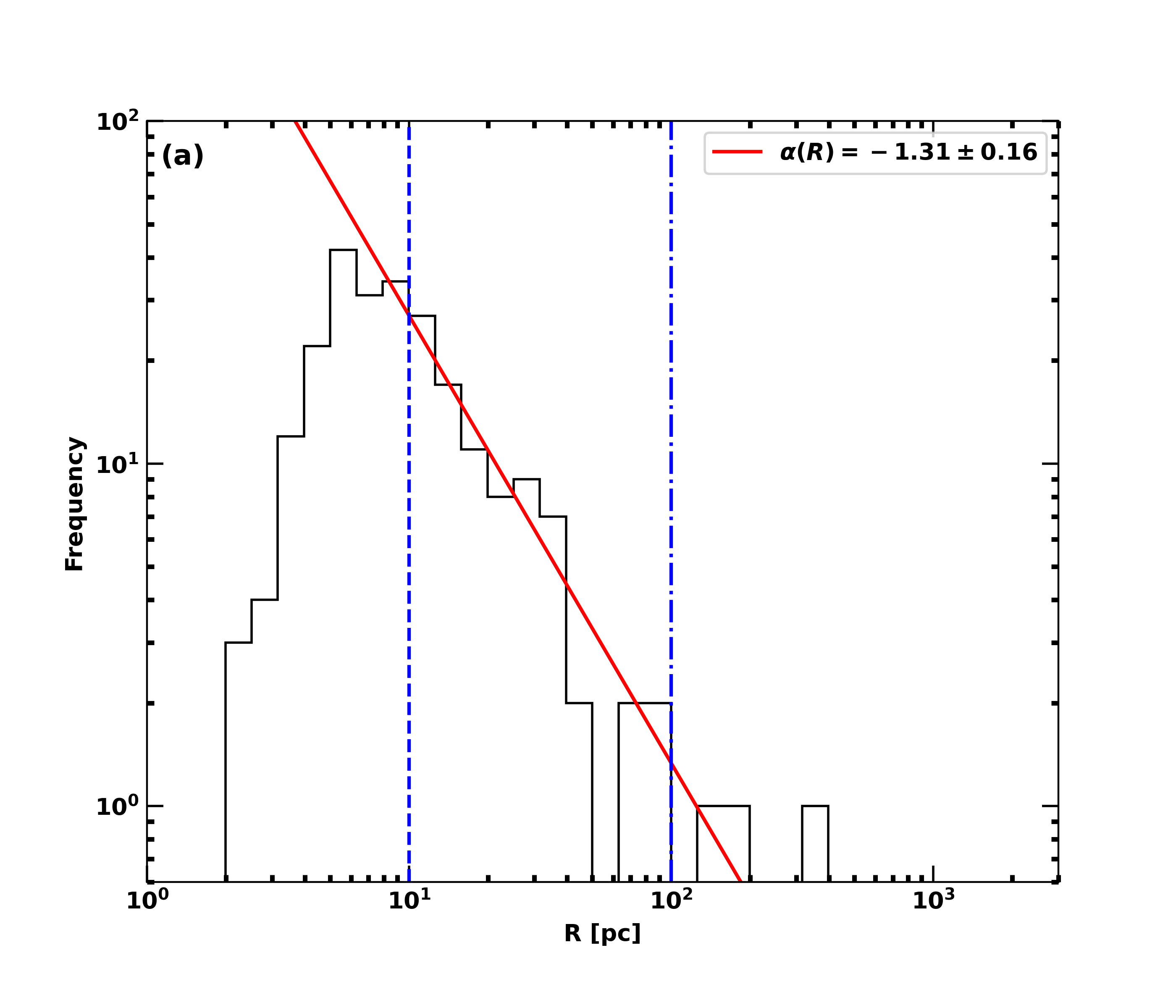}\\
    \includegraphics[width=\columnwidth,trim={0.6cm 0.2cm 1.5cm 1.2cm},clip]{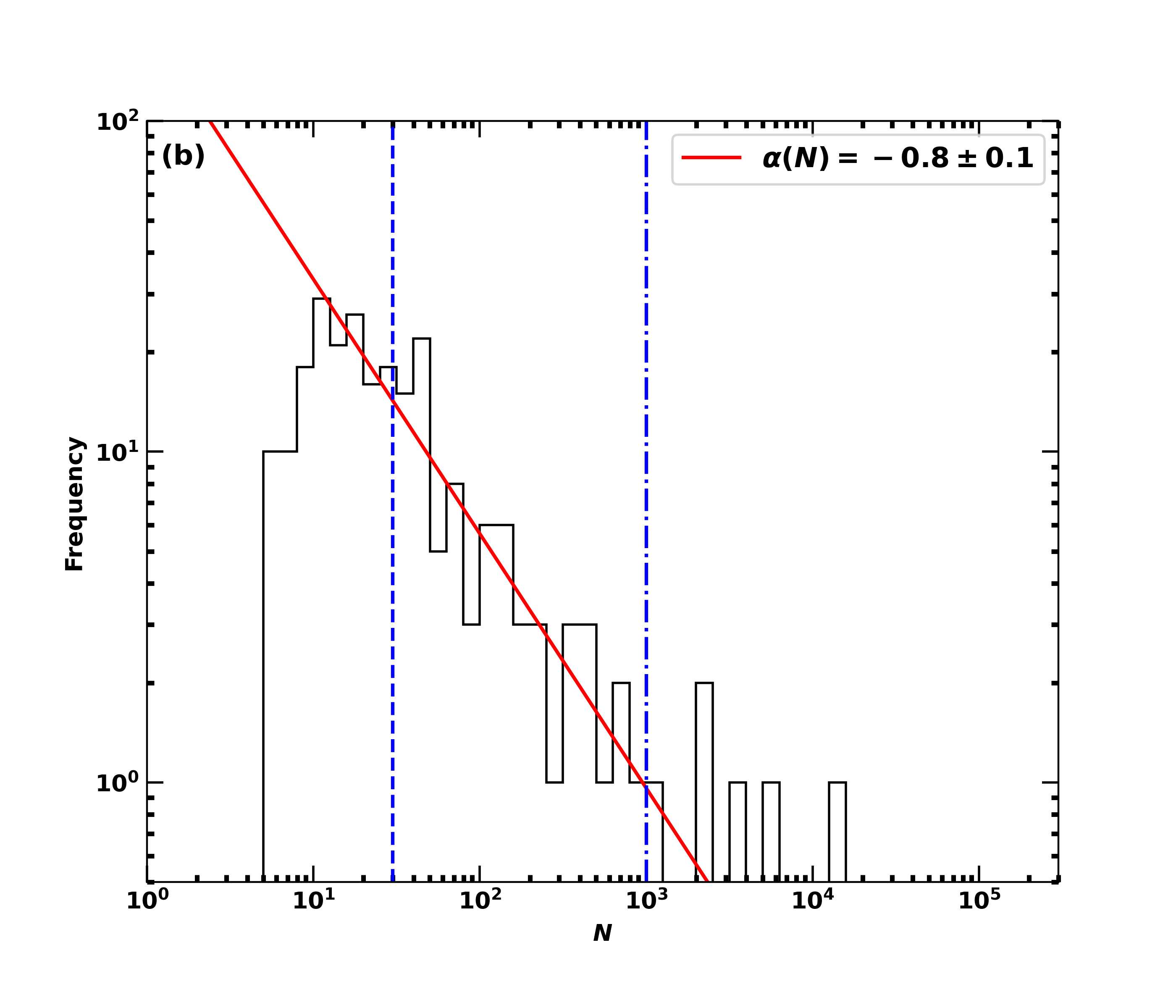}\\
    \includegraphics[width=\columnwidth,trim={0.6cm 0.2cm 1.5cm 1.2cm},clip]{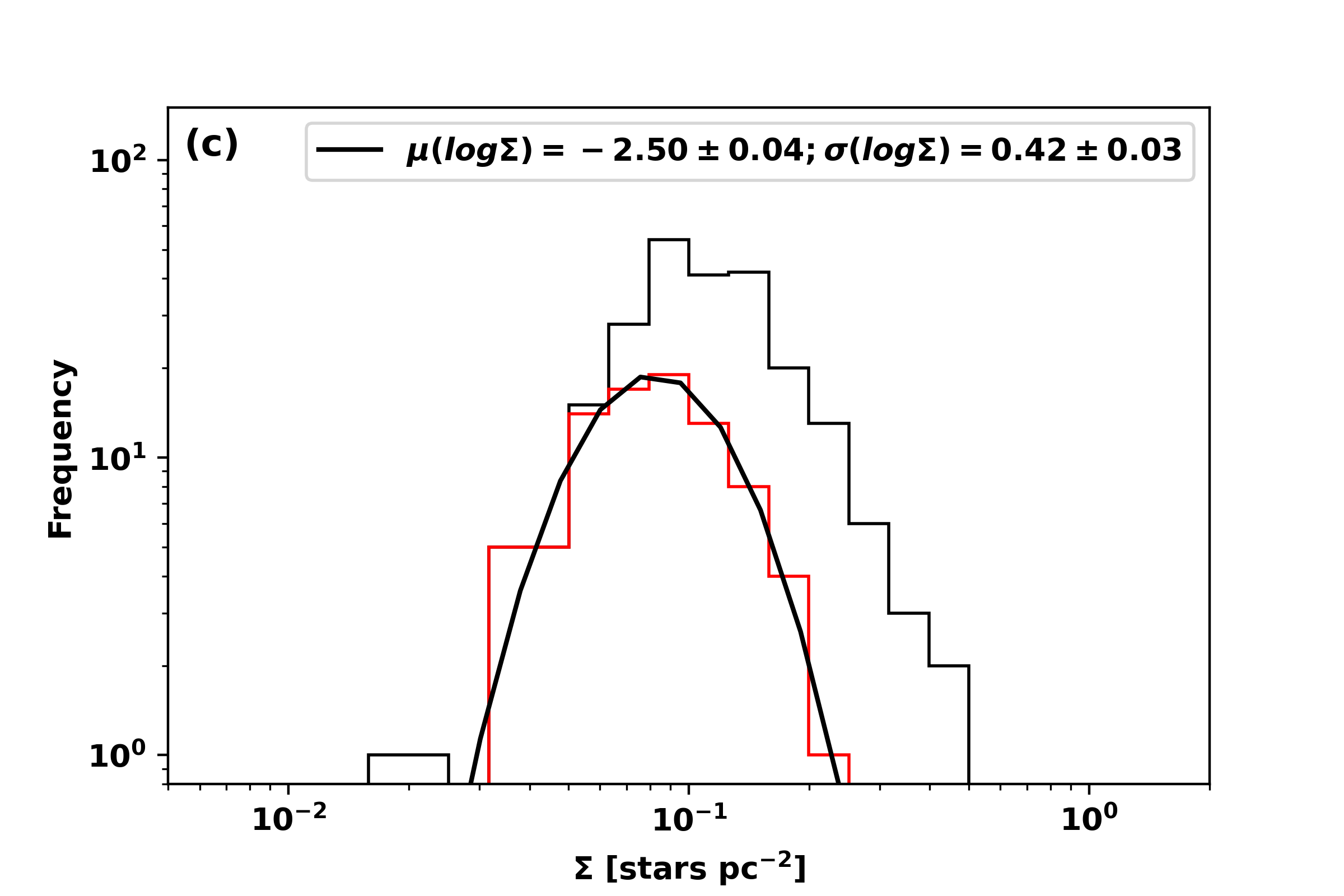}
    \caption{Distributions of (a) size, (b) number, and (c) surface density for the identified young stellar structures. In panels (a) and (b), the blue dashed lines indicate the range of data points used for power-law fitting, with the slope of the fit denoted by $\alpha$. Panel (c) displays the surface density distribution, where the black histogram includes all data points, and the red histogram represents the distribution after applying constraints.}
    \label{fig:mass_size_surface_density_dist}
\end{figure}

\autoref{fig:mass_size_surface_density_dist}(a) shows the distribution of the size of the detected young stellar structures. We note that the peak, mean, and median of the size distribution are at 5.6 pc, 14.6 pc, and 8.2 pc, respectively. As noted in Section~\ref{subsec:kde}, structures with sizes smaller than 10 pc are not resolved, and statistical noise is evident in the size distribution, particularly for larger sizes. Therefore, to fit a single power law, we selected the range from 10 pc to 100 pc in the size distribution of the detected young stellar structures. We performed a fit to the data within the range 10--100 pc, as shown by the red solid line in \autoref{fig:mass_size_surface_density_dist}(a), and estimated the best-fitting slope, $\alpha (R) = -1.31 \pm 0.16$. Substructures within an hierarchical system follow a cumulative size distribution,
\begin{equation}
    N(>R) \propto R^{-D_2},\label{eq:first}
\end{equation}
where $D_2$ is the 2D fractal dimension \citep{1983fractal,2017...Sun..30,2017..Sun..LMC..Bar,2018...Sun..SMC,2022MNRAS..Miller..LMC}. Since Equation~\ref{eq:first} represents a single power law, it is mathematically equivalent to a differential size distribution,
\begin{equation}
    \frac{{\rm d}N}{{\rm d} \log R} \propto R^{-D_2}.\label{eq:second}
\end{equation}
Equation~\ref{eq:second} follows the same form as the size distribution of the detected young stellar structures (\autoref{fig:mass_size_surface_density_dist}a) for a size range of 10--100 pc. Consequently, these structures align with a 2D fractal dimension of $D_2 = -\alpha(R) = 1.31 \pm 0.16$, similar within 2$\sigma$ to the $D_2$ value derived from the number--size relation (i.e. $1.64 \pm 0.03$, as discussed in Section~\ref{subsec:MSR}). We note that there are three structures with sizes of $100 < R < 400$ pc. We found their linear scales to be comparable to that of the main body of the SMC, consistent with \citet{2018...Sun..SMC}. This suggests that the spatial extent of the larger structures composed of young massive stars (5–40 $M_\odot$) identified in our study is similar to that of large-scale structures traced by low-mass young stellar populations \citep{2018...Sun..SMC}.Structures located in the inner region of the SMC are less affected by external galactic processes---such as tidal forces and interactions with neighboring galaxies---compared with those in the outermost regions \citep{2020..De..Leo..tidal..scars..SMC,2022..Dias..tidal..radius}. Hence, the larger, outer structures are more affected by such global processes, which likely contributes to their deviation from the power-law size distribution.

\autoref{fig:mass_size_surface_density_dist}(b) shows the number distribution of the detected young stellar structures. Here, we confirm that the number distribution is incomplete for smaller values (i.e. smaller numbers of stars). Based on \autoref{fig:mass_size_rel}, we note that for $N \gtrsim 30$, the sizes correspond to $R > 10$ pc, i.e. there is a negligible effect of incompleteness in detecting young stellar structures for $N \gtrsim 30$. For $N > 1000$, we note the increased statistical noise in the number distribution. That is the reason why we have fitted only a single power law in the range encompassing 30--1000 in the number distribution of the detected structures. If masses are arranged in an hierarchical pattern, their mass distribution will follow a single power law with a slope of $-1$ \citep{1996..Elmegreen..HSF,2018...Sun..SMC,2022MNRAS..Miller..LMC}. Similarly as in Section~\ref{subsec:MSR}, the slope of the number distribution can serve as a reliable proxy for the slope of the mass distribution. From the best fits, we found a power-law slope of $\alpha(N) = -0.8 \pm 0.1$, which is close to the expected theoretical value. Five structures have $N > 1000$; they are not well-represented by the same power-law slope. Note that the two structures with the highest and second-highest $N$ values have projected boundaries comparable in extent with that of the main body of the SMC, while the remaining three structures have sizes of nearly 100 pc. These two high-$N$ structures are, therefore, more likely associated with galactic-scale processes rather than with sub-galactic hierarchical star formation \citep{2020..De..Leo..tidal..scars..SMC,2022..Dias..tidal..radius}.

\autoref{fig:mass_size_surface_density_dist}(c) shows the distribution of the surface density of all detected young stellar structures, represented by the black histogram. Note that there are two structures with very low surface densities (below 0.03 pc$^{-2}$), which introduces statistical noise into the distribution. For this analysis, therefore, these two structures, along with those having a size of $R \leq 10$ pc, have been excluded. The distribution of the surface density for the remaining structures is shown as the red histogram in \autoref{fig:mass_size_surface_density_dist}(c). The structures considered follow a log-normal distribution in surface density, as shown by the solid black curve. The mean and standard deviation of the logarithm of the surface density are $\mu (\log\Sigma) = -2.50 \pm 0.04$ (stars $pc^{-2}$) and $\sigma (\log\Sigma) = 0.42 \pm 0.03$ (stars $pc^{-2}$).


\section{Discussion}
\label{sec:dis}

\subsection{Impact of different constraints on the results}
\label{subsec:diff_cons}
\begin{figure*}[ht]
    \centering
    \includegraphics[width=\linewidth]{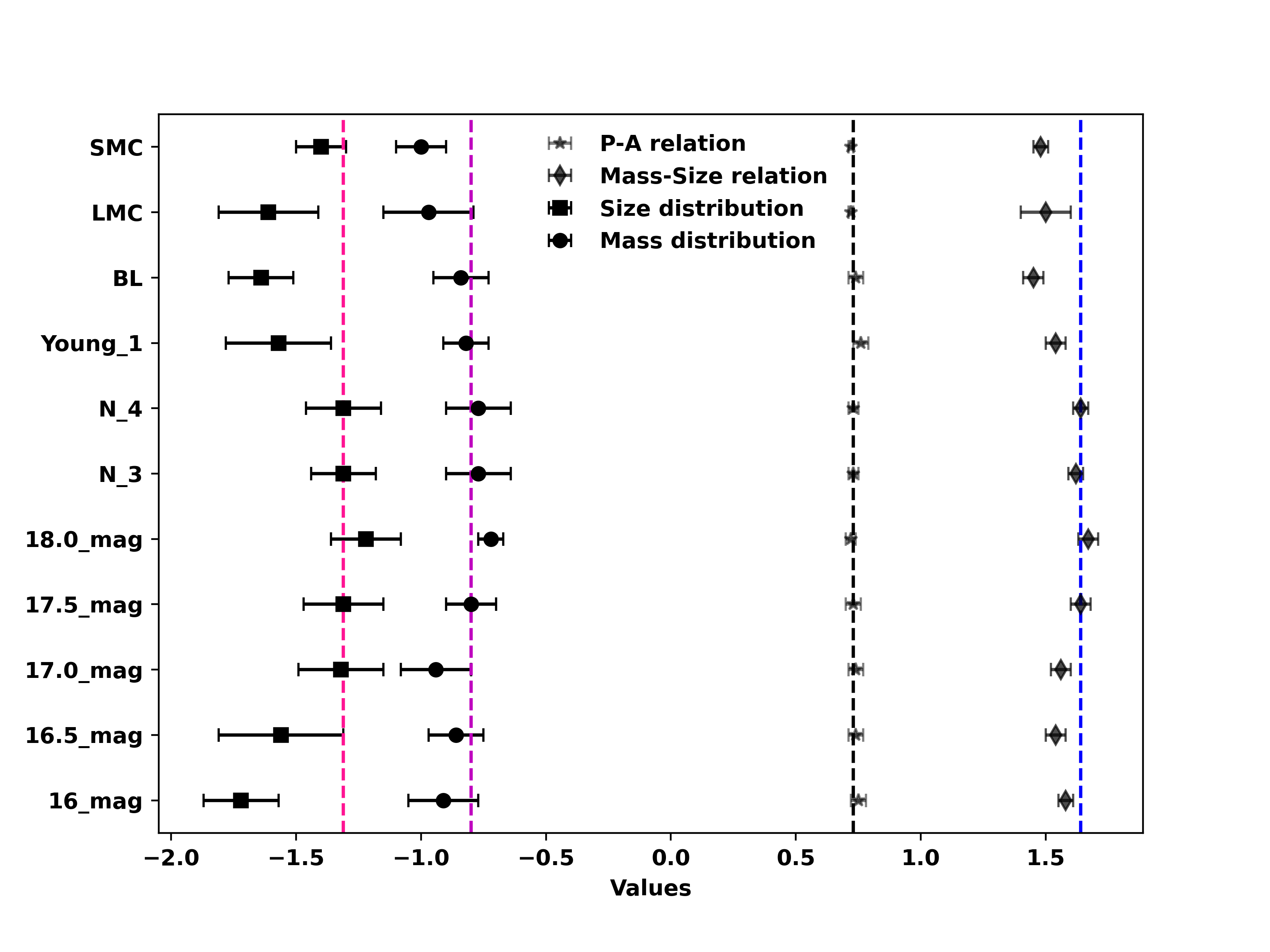}
    \caption{Comparison of the power-law slopes derived from the distributions of the number and size, perimeter--area, and number--size relationships, under varying constraints for $N_{\rm min}$, magnitude cut-off, different populations, and literature results for the LMC and SMC. The labels `LMC' and `SMC' along the $Y$ axis are associated with values from the literature; see \citet{2022MNRAS..Miller..LMC} and \citet{2018...Sun..SMC}, respectively. The dashed lines represent the slopes obtained for the sample in this study with an FUV magnitude cutoff of 17.5 and $N_{\rm min}=5$.}
    \label{fig:dis_plot}
\end{figure*}

In this study, we compare the outcomes under varying conditions, such as for different magnitude cut-offs (which affect data completeness), populations characterized by different ages, and different number cut-offs for structure detection, with the results obtained for the sample with an FUV magnitude cut-off of 17.5 and $N_{\rm min}$. \autoref{fig:dis_plot} shows the slopes of the power-law relations obtained via  least-squares fitting of the number distribution, size distributions, the perimeter--area and the number--size relationships, as detailed in Sections~\ref{subsec:PAR}, \ref{subsec:MSR}, and \ref{subsec:DSMSD}, under varying conditions. It also includes literature values for the LMC \citep{2022MNRAS..Miller..LMC} and the SMC \citep{2018...Sun..SMC}, as discussed in Section~\ref{subsec:cwow}. The dashed lines indicate the slopes derived for the sample in this study, using an FUV magnitude cut-off of 17.5 and a minimum count of $N_{\rm min}=5$.

We first examined the impact of incompleteness by applying different magnitude cut-offs,\footnote{Each FUV magnitude cut-off corresponds to a cumulative age threshold, with brighter limits selecting progressively younger populations. These cut-offs were primarily used to assess the effects of photometric incompleteness on the structure identification process.} ranging from 16 to 18 mag in steps of 0.5 mag, while retaining a constant number cut-off ($N_{\rm min} = 5$) for identifying young stellar structures. As shown in \autoref{fig:dis_plot}, the slopes derived from the perimeter--area relations remain consistent within the uncertainties, and the slopes obtained from the size distributions, number distributions, and number--size relations all fall within the 2$\sigma$ range of the slopes for the entire sample. These findings indicate that within a data completeness range of 90--100\%, the hierarchical nature of the young stellar structures remains unaffected. Since our results remain the same for different magnitude cut-offs, i.e., at 17, 17.5, and 18 mag, it follows that depth and extinction effects are negligible for our sample (younger than 150 Myr).

To eliminate spurious detections of young stellar structures, we established a minimum cut-off for the number of stars, denoted by $N_{\rm min}$. A structure is considered genuine if it contains at least $N_{\rm min}$ stars. We set $N_{\rm min} = 5$. To explore the effects of a lower $N_{\rm min}$ value, which could introduce more noise, we also examined results with cut-offs of $N_{\rm min} = 3$ and 4, while keeping a magnitude cut-off of 17.5 mag. Our results, shown in \autoref{fig:dis_plot}, demonstrate that the slopes of various relations and distributions remain consistent, indicating that $N_{\rm min} = 3$ is acceptable, since the effect of noise on the final results is negligible.

To examine how the fractal nature of young populations evolves with age, we analyzed two young stellar populations (Young 1 and Blue Loop; BL) spanning different age groups, identified from the $G$ vs. $(G_{\rm BP} - G_{\rm RP})$ CMDs of \citet{2021..Gaia} and H24 (their Figure 5b). Here, we did not consider the Young 2 and Young 3 populations, given that the majority of those stars are fainter than 17.5 mag. We note that the BL (younger than 200 Myr) and Young 1 (age $\leq$ 50 Myr) populations include some stars with FUV magnitudes fainter than 18 mag; however, we have presented results for them without accounting for incompleteness effects. The derived slopes from the perimeter--area relation and number distribution fall within the uncertainties of our earlier results, as shown in \autoref{fig:dis_plot}. Additionally, the fractal dimensions of the hierarchical structures derived from the number--size relation and the size distribution remain within the 2$\sigma$ range. The Young 1 and BL stars show irregular morphologies and fractal characteristics. This indicates that the fractal nature of star formation remains present in the SMC for populations with mean ages as old as 200 Myr.

\subsection{Similarities Between the ISM and Young stellar structures}
\label{subsec:SBIYSS}

The young stellar structures and the ISM exhibit many notable similarities. The gas exhibits a hierarchical, self-similar morphology across a broad range of scales, from 0.1 pc to 1 kpc \citep{2004..Elmegreen..Scalo,2007..Bergin..Tafalla,2008..Sanchez}. The largest gas structures are comparable in size with the parent galaxy itself \citep{2000..Elmegreen}. In our case, the sizes of the young stellar structures range from a few parsecs to several hundred parsecs (see Section~\ref{sec:CDP} and \autoref{fig:mass_size_rel}). The gas substructures exhibit irregular morphologies. The projected boundaries of these structures have been studied extensively using perimeter--area relations, including by \citet{1987..Beech..Molcld,1991..Falgarone..MolCld,1994..Vogelaar..MolCld,2016..Lee..Molcld}. The perimeter--area dimension ($D_p$) typically falls between 1.3 and 1.6 \citep{1990..Dickman,1993..Hetem,2010..Julia}, and the value derived here, $D_p = 1.46 \pm 0.04$ (see Section~\ref{subsec:PAR}), also lies within this range.\footnote{The $D_p$ values obtained under different conditions, as discussed in Section~\ref{subsec:diff_cons}, also fall within the range found in the literature.}

Turbulence significantly affects the hierarchical substructures within the ISM, acting as a key driver for the fragmentation of the ISM into increasingly smaller substructures and the formation of stars \citep{2004..Elmegreen..Scalo,2018..Pingel..turbulence}. Turbulence forms across a wide range of spatial scales and cascades down to the smallest scales, as evidenced by the fractal nature of the ISM \citep{2001..Elmegreen..HSF..LMC}. On larger scales, turbulence is likely initiated by the accretion of circumgalactic material and gravitational instabilities \citep{2016..Krumholz..Turbulence}, whereas on smaller scales, energy is likely injected by stellar feedback mechanisms such as outflows and supernova explosions \citep{2016..Padoan..SN}. The H{\sc i} gas in both Magellanic Clouds exhibits a fractal structure on galactic scales \citep{1999..SMC..Stanimirovic..SMC..GLXstr,2003..Kim..LMC..GLXstr,2017..Nestingen}. \citet{2001..Stanimirovic..SMC..HI} analyzed H{\sc i} data from the Australia Telescope Compact Array and the Parkes Telescope/Murriyang, examining the spectrum of intensity fluctuations in relation to the velocity slice thickness. Their results were consistent with the theoretical predictions of turbulence in the SMC by \citet{2000..Lazarian..Pogosyan}. Thus, the pronounced density fluctuations of H{\sc i} in the SMC result from active turbulence. \citet{1999..SMC..Stanimirovic..SMC..GLXstr,2000..Stanimirovic} estimated projected fractal dimensions of $D_2 = 1.4$--1.5 for the gas and dust in the SMC. These values are consistent with the fractal dimensions of our detected young stellar structures: $D_2 = 1.64 \pm 0.03$ from the number--size relation and $D_2 = 1.31 \pm 0.16$ from the size distribution (see \autoref{fig:mass_size_rel} and \autoref{fig:mass_size_surface_density_dist}a). This study, therefore, supports a scenario of hierarchical star formation within a turbulent ISM, where the properties of the ISM are imprinted onto the youngest stellar structures across the scale. The striking similarities between the ISM and these stellar structures further indicate that dynamical evolutionary effects between the two are likely minimal. Hierarchical structures result from both top-down (fragmentation of molecular clouds) and bottom-up (energy injection from stellar winds and supernovae) mechanisms \citep{2022MNRAS..Miller..LMC}. Smaller structures reflect the fractal nature of molecular clouds, while larger ones align with the large-scale ISM shaped by supersonic turbulence.

The impact of turbulence aligns with the observed surface density distribution. Studies of molecular clouds reveal that those exhibiting a log-normal surface density distribution are primarily governed by supersonic turbulence \citep{2002..Padoan..gas..LN,2011..Elmegreen..gas..LN,2017..Gouliermis..HSF..CT}. Log-normal distributions have been identified in the column and volume densities of molecular clouds \citep{2010..Lombardi..Gas..LN,2012..Konstandin..gas..LN} as well as in simulations of turbulent gas \citep{2000..Klessen..LN,2010..Federrath..gas..LN,2012..Konstandin..gas..LN}. Since star formation follows the gas distribution, young stellar structures are expected to exhibit a log-normal surface density distribution, consistent with our results discussed in Section~\ref{subsec:DSMSD} and shown in \autoref{fig:mass_size_surface_density_dist}c. In contrast, molecular clouds with a surface density distribution that exhibits a power-law tail are more strongly affected by self-gravity \citep{2000..Klessen..LN, 2010..Federrath..gas..LN}. However, note that the lack of a power-law tail in our size distribution may be caused by the limitations of our study, specifically by the minimum spatial resolution of our KDE map of 10 pc. This resolution limit might prevent us from detecting the effects of self-gravity, given that we are unable to resolve the smallest and densest structures \citep{2018...Sun..SMC,2022MNRAS..Miller..LMC}.

\subsection{Hierarchy in galaxies}
\label{subsec:cwow}

Our analysis has shown that hierarchical substructures among young stellar populations in the SMC are preserved up to ages of $\sim$200 Myr (Section~\ref{subsec:diff_cons}), thus clearly extending the previously reported upper limit of $\sim$75 Myr by \citet{2008..Gieles..Stellar..Str..SMC}. This suggests a more prolonged dispersal timescale of stellar structures in the SMC than previously inferred. In our study, the size distribution yields a fractal dimension of $D_2 = 1.31 \pm 0.16$. This is consistent with the values reported for young clusters in the Magellanic Clouds by \citet{2010..Bonatto..HSF..L&SMC}. \citet{2014..Gouuliermis..NGC346} derived a fractal dimension of $D_2 = 1.2$--$1.4$ for NGC 346, a local star-forming region within the SMC, based on simulations, which is broadly consistent with our global estimate for the entire SMC.

\citet{2018...Sun..SMC} used a contour-based method to examine the fractal nature of young stellar structures in the SMC using VMC data. They identified 556 young stellar structures, whereas our analysis detected roughly half that number. This difference is attributed to the larger coverage area of the SMC in their VMC-based study. Despite differences in coverage, the slopes of the perimeter--area relation and the number distribution are consistent within the uncertainties, and the fractal dimension ($D_2$) falls within the 1.5$\sigma$ range of the sample, as shown in \autoref{fig:dis_plot}.

The fractal dimensions derived in our study, as well as that of \citet{2018...Sun..SMC} for the SMC, are smaller than the value of $D_2 = 1.8$ reported by \citet{2009..Bastian..HSF.LMC} for the LMC. The latter authors used fractured minimum spanning trees, the $Q$ parameter, and two-point correlation functions, methods known to involve considerable uncertainties \citep{2004..Cartwright..Whitworth}. In contrast, our results are consistent within the prevailing uncertainties with the fractal dimensions of the LMC Bar ($D_2 = 1.5 \pm 0.1$) and 30 Doradus ($D_2 = 1.6 \pm 0.3$) obtained using similar contour-based techniques \citep{2017..Sun..LMC..Bar, 2017...Sun..30}. When we compare our findings with those obtained for the entire LMC by \citet[VMC data and contour-based analysis;][]{2022MNRAS..Miller..LMC}, we observe that the young stellar structures in the LMC exhibit comparable power-law relations and distributions to those identified in the SMC (see \autoref{fig:dis_plot}). More recently, \citet[][; A. Miller et al., submitted]{2024..Miller..Automated..code} developed a novel automated technique for detecting and characterizing semi-resolved star clusters, where the point-spread function (PSF) is smaller than the cluster size but larger than typical stellar separations. This method is well-suited for studying resolved structures in nearby galaxies. Applying it to a 1.77 deg$^2$ area in the LMC, they identified 682 cluster candidates. Overall, the comparison of fractal dimensions and parameter distributions indicates that, despite differences in galactic environments---such as metallicity and the spatial extent of the surveyed regions (e.g., 30 Dor, LMC Bar, entire LMC and SMC)---young stellar structures in both the LMC and SMC consistently display an hierarchical spatial pattern.

\citet{2020..SF..regions..MW} estimated the three-dimensional fractal dimension ($D_3 = D_2 + 1$) for three Milky Way star-forming regions using the box-counting method. They found $D_3$ values of 2.468 for M16 ($D_2 = 1.468$), 2.126 for the Orion Nebula ($D_2 = 1.126$), and 2.435 for RCW 38 ($D_2 = 1.435$). These values are consistent with the uncertainties in our derived $D_2$ measurements. Additionally, our $D_2$ value aligns well with that of open clusters younger than 100 Myr in the Milky Way \citep{2025..Fractal..dimension..open..clusters..MW}. This suggests that galaxies of different ordered masses, such as the Milky Way, LMC, and SMC, exhibit hierarchical star formation characterized by a similar fractal dimension.

In this paper, the sizes of the young stellar structures in the SMC range from a few parsecs to several hundred parsecs. Similar size ranges have been observed in six nearby (3--15 Mpc) star-forming galaxies (50--150 pc; \citealt{2017..Grasha..6..glx}), IC 2574 (15--285 pc; \citealt{2019..Mondal..IC2574}), NGC 7793 (12--70 pc; \citealt{2021..Mondal}), the Wolf--Lundmark--Melotte galaxy (7--30 pc; \citealt{2021..Mondal..b}), and the LMC (10--700 pc; \citealt{2022MNRAS..Miller..LMC}). 

By comparing findings from the MW, LMC, SMC, and other galaxies discussed in this section, we assess whether the hierarchical properties of star formation are influenced by factors such as metallicity or environmental conditions (e.g., starburst activity, galaxy interactions). The results suggest that there is no clear dependence on any of these parameters. This indicates that, irrespective of large-scale galactic dynamics or small-scale microphysical processes, the clustering of star formation across scales from stellar clusters to entire galaxies appears to be governed universally by turbulence-driven hierarchical star formation.

\section{Conclusion and Summary}
\label{sec:cs}

We present the first FUV study of hierarchical star formation in the SMC to date. Our results and conclusions can be summarized as follows:

\begin{itemize}
    \item We have detected 236 young stellar structures composed of FUV stars with magnitudes brighter than 17.5 mag and a completeness level greater than 90\%.
    \item We have calculated the key parameters of the detected young stellar structures, such as their sizes, areas, perimeters, surface densities, and the number of stars within a given structure.
    \item The sizes of the structures vary from a few parsecs to a few hundred pc. 
    \item From the perimeter--area relation, the estimated perimeter--area dimension is $D_p = 1.46 \pm 0.04$.
    \item The 2D fractal dimensions obtained from the number--size relation and the size distributions are $D_2 = 1.64 \pm 0.03$ and $D_2 = 1.31 \pm 0.16$, respectively. These two values are equivalent within their 2$\sigma$ errors.
    \item From the best fit to the number distribution between 30 and 1000, we estimated a slope of $-0.8 \pm 0.1$, which is close to the theoretically expected value of $-1$. 
    \item The surface density distribution of the detected young structures follows a log-normal distribution.
    \item We have tested our results under various conditions, including the imposition of a magnitude cut-off, a number cut-off, and age limits, and found that all results remain consistent within the 2$\sigma$ range of the values obtained for the sample. 
    \item We trace the fractal nature of star formation in populations up to an age of 200 Myr.
\end{itemize}

Our results for the stellar distributions and fractal dimensions $D_p$ and $D_2$ are consistent with those observed in the SMC, LMC, MW and other galaxies, highlighting strong similarities in both the fractal dimensions and the surface density distributions of stellar structures. These characteristics mirror those seen in the ISM based on both observations and numerical simulations, suggesting that the stellar structures we identified have inherited their fractal nature from the ISM. This supports the idea that supersonic turbulence plays a key role in shaping the hierarchical architecture of these structures. Our results align with the broader theory of turbulence-driven hierarchical star formation observed within different galactic environments.

\section*{acknowledgement}
We want to thank the anonymous referee for the insightful review that helped in the improvement of the paper. 
AS acknowledges support from SERB for her POWER fellowship. SH also acknowledges funding from the Australia--India Research Studentship (AIRS) program. This publication uses the catalog compiled by \citep{2024Hota..FUV..catalog..SMC} based on data from the AstroSat mission of the Indian Space Research Organization, archived at the Indian Space Science Data Centre (ISSDC). 

Software: PYTHON packages, like NUMPY \citep{2020Natur.585..357H}, ASTROPY \citep{2013A&A...558A..33A,2018..Astropy,2022..Astropy}, MATPLOTLIB \citep{2007CSE.....9...90H} and SCIPY \citep{2020NatMe..17..261V}.

\end{document}